\documentclass[12pt]{article}
\usepackage{graphicx}

\newcommand\pubnumber{IPPP/14/85}
\newcommand\pubdate{\today}

\def\napoli{IPPP Durham}

\def\Title#1{\begin{center} {\Large #1 } \end{center}}
\def\Author#1{\begin{center}{ \sc #1} \end{center}}
\def\Address#1{\begin{center}{ \it #1} \end{center}}

\newcommand\pubblock{\rightline{\begin{tabular}{l} \pubnumber\\
         \pubdate  \end{tabular}}}
\newenvironment{Abstract}{\begin{quotation}  }{\end{quotation}}
\newenvironment{Presented}{\begin{quotation} \begin{center} 
             PRESENTED AT\end{center}\bigskip 
      \begin{center}\begin{large}}{\end{large}\end{center} \end{quotation}}
\def\Acknowledgements{\bigskip  \bigskip \begin{center} \begin{large}
             \bf ACKNOWLEDGEMENTS \end{large}\end{center}}

\def\beq{\begin{equation}}
\def\eeq#1{\label{#1}\end{equation}}
\def\eeqn{\end{equation}}

\def\beqa{\begin{eqnarray}}
\def\eeqa#1{\label{#1}\end{eqnarray}}
\def\eeqan{\end{eqnarray}}

\let\bar=\overbar

\def\Dslash{\not{\hbox{\kern-4pt $D$}}}
\def\dslash{\not{\hbox{\kern-2pt $\del$}}}

\def\msb{{\bar{\ssstyle M \kern -1pt S}}}

\begin{document}
\begin{titlepage}
\pubblock

\vfill
\Title{$B$-mixing in and beyond the Standard model}
\vfill
\Author{Alexander Lenz}
\Address{\napoli}
\vfill
\begin{Abstract}
We review the status of mixing of neutral $B$-mesons, 
including a discussion of the current precision of 
Standard Model (SM) predictions as well as the space that is left for effects
of new physics. In that respect we present several observables, which
are particularly sensitive to the remaining new physics (NP) parameter space.
$B$-mixing can also be used to test the fundaments of quantum mechanics,
here we suggest a new measurement of the ratio of like-sign dilepton 
events to opposite-sign dilepton events. Finally we summarise
briefly the status of lifetimes of heavy hadrons. The corresponding 
theory predictions rely on the same tool - the Heavy Quark Expansion (HQE) 
- as some of the mixing quantities. New experimental data has recently
proven the validity of the HQE to a high accuracy. However, the theoretical precision  of lifetime predictions is strongly limited by a lack of 
non-perturbative evaluations of matrix elements of dimension-six operators.
\end{Abstract}
\vfill
\begin{Presented}
8th International Workshop on the CKM Unitarity Traingle\\
8-12 September 2014, Vienna, Austria
\end{Presented}
\vfill
\end{titlepage}
\def\thefootnote{\fnsymbol{footnote}}
\setcounter{footnote}{0}

\section{Introduction}
\begin{figure}
\includegraphics[width=0.9\textwidth,angle = 0]{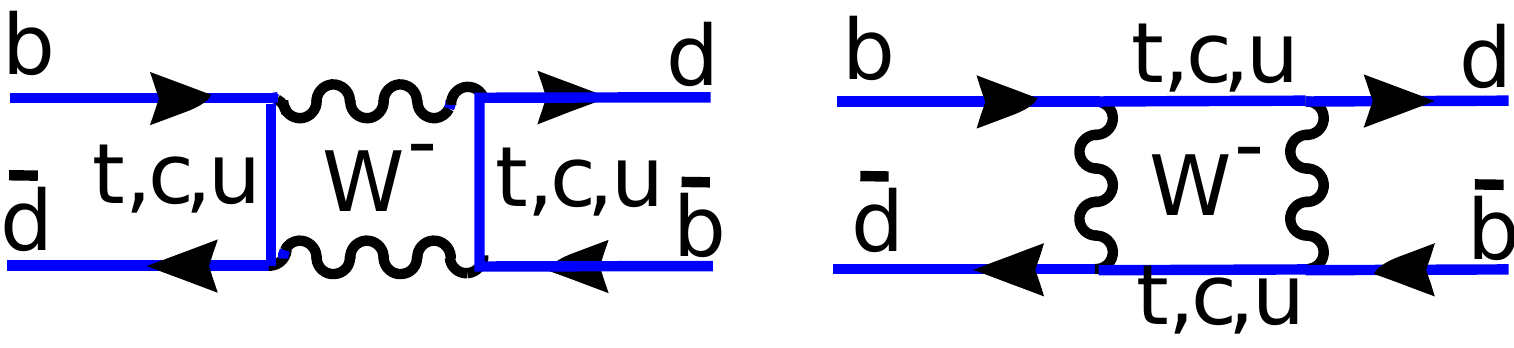}
\caption{Box diagrams triggering the transition of a $\bar{B}_d$-meson into a $B_d$ meson.}
\label{box}
\end{figure}

Mixing of neutral mesons is a macroscopic quantum effect that is triggered
by the so-called box diagrams shown in Fig. \ref{box}, see e.g. the reviews
\cite{Anikeev:2001rk,Bediaga:2012py,Lenz:2014nka,Lenz:2012mb} for a more detailed discussion 
and also some historical remarks. In the SM these transitions are suppressed
by being a second order weak interaction process. NP contributions to mixing
thus might be easily of a similar size as the SM contribution.
Calculating the on-shell part of the box diagrams gives 
$\Gamma_{12}^q$ ($q = d,s$) and the off-shell part gives 
$M_{12}^q$. Because of the CKM structure both $\Gamma_{12}^q$ and 
$M_{12}^q$ can be complex. The three quantities
$ |M_{12}^q|$, $ |\Gamma_{12}^q|$ and 
$ \phi_q = \mbox{arg}( -M_{12}/\Gamma_{12})$
can be related to three observables:
\begin{enumerate}
\item The mass difference of the two mass eigenstates $B_H$ and $B_L$: 
      \begin{equation}
      \Delta M_q := M_H - M_L  \approx 
        2 |M_{12}^q|  \; .
      \end{equation}
      As $M_{12}^q$ is given by the off-shell intermediate states, it is sensitive to
      heavy internal particles. In the SM these are the $W$-boson and the top-quark;
      depending on your favourite model for NP, these might also be e.g. heavy  SUSY-particles,
      see e.g. \cite{kubo}.
      Hence $\Delta M_q$ is supposed to be sensitive to NP effects originating at a high scale.
\item The decay rate difference of the two mass eigenstates $B_H$ and $B_L$:
       \begin{equation}
       \Delta \Gamma_q := \Gamma_L - \Gamma_H \approx
        2 |\Gamma_{12}^q| \cos  \phi_q  \; .
       \end{equation}
       As $\Gamma_{12}^q$ is given by on-shell intermediate states, it is sensitive to 
      light internal particles, like the up- and charm-quark in the SM. At first sight it seems
      reasonable to assume almost no NP effects in $\Gamma_{12}^q$ - later on we will 
      challenge this assumption. $\Delta \Gamma_q$ can of course always be affected by new physics 
      effects in the phase $\phi_q$.
\item Flavour specific (or more specific semi-leptonic) CP asymmetries can also be 
      expressed in terms of the three mixing quantities $\Gamma_{12}^q$,  $M_{12}^q$ and $\phi_q$.
      \begin{eqnarray}
        a_{sl}^q & \equiv & a_{fs}^q = 
       \frac{\Gamma \left(\bar{B}_q(t) \to f\right)-\Gamma \left({B}_q(t) \to \bar{f} \right)}
            {\Gamma \left(\bar{B}_q(t) \to f\right)+\Gamma \left({B}_q(t) \to \bar{f} \right)}
      = \left| \frac{\Gamma_{12}^q}{M_{12}^q} \right| \sin \phi_q \; .
      \end{eqnarray}
      Since both $\Gamma_{12}/M_{12}^q$ and $\phi_q$ are small in the SM, the semi-leptonic
      CP asymmetries provide a powerful null test.
\end{enumerate}

\section{Standard model predictions}
\subsection{Mass difference}
The SM expression for $M_{12}^q$ is given as
        \begin{eqnarray}
        M_{12,q} & = & \frac{G_F^2}{12 \pi^2} 
          (V_{tq}^* V_{tb})^2 M_W^2 S_0(x_t)
          { B_{B_q} f_{B_q}^2  M_{B_q}} \hat{\eta }_B \; .
        \end{eqnarray}
The 1-loop result for the box-diagram is denoted by the Inami-Lim 
function $S_0(x_t)$ \cite{Inami:1980fz}, NLO-QCD corrections to the 
box-diagrams  by $\hat{\eta }_B$ \cite{Buras:1990fn} and 
non-perturbative contributions by the bag parameter 
$B_{B_q}$ and the decay constant $f_{B_q}$.
Taking the FLAG-average \cite{Aoki:2013ldr} for $f_{B_q}^2 B_{B_q}$
we obtain the SM prediction, which can be compared to the experimental
averages given by HFAG \cite{HFAG}: 
\begin{eqnarray}
\Delta M_d^{\rm SM}  = 0.543  \pm 0.091 \; \mbox{ps}^{-1}\; , & & 
\Delta M_d^{\rm Exp} = 0.510  \pm 0.003 \; \mbox{ps}^{-1} \; ,
\\
\Delta M_s^{\rm SM}  = 17.30  \pm 2.6   \; \mbox{ps}^{-1} \; ,& & 
\Delta M_s^{\rm Exp} = 17.761 \pm 0.022  \; \mbox{ps}^{-1} \; .
\end{eqnarray}
The measurements agree very nicely with the SM predictions, but the theoretical
uncertainties are considerably larger than the experimental ones. Thus we still
have quite some space for NP effects. The theoretical error is dominated by the
non-perturbative uncertainties in $B_{B_q}$ and $f_{B_q}$. Also some of the lattice
predictions yield quite different values; compare e.g. the  determinations
from Fermilab/MILC \cite{Bazavov:2011aa} and the one from HPQCD \cite{Dowdall:2013tga}:
\begin{eqnarray}
f_{B_s}^{\rm Fermilab/MILC} = 242.0 \pm 5.1 \pm  8.0 \; {\rm Mev} \; ,
&&
f_{B_s}^{\rm HPQCD} = 224   \pm 5  \; {\rm Mev} \; .
\end{eqnarray}
In view of the quadratic dependence of many observables on the decay constant
further lattice studies would be very helpful.

\subsection{Heavy Quark Expansion}
The theoretical prediction of $\Gamma_{12}^q$ is more involved than the one of $M_{12}^q$, 
here a second operator product expansion has to be performed, the so-called
Heavy Quark Expansion (HQE), see e.g. \cite{Lenz:2014jha} for a review of this theoretical tool.
The HQE applies also for lifetimes and totally inclusive decays decay rates of  heavy hadrons.
Historically there had been several discrepancies between experiment and theory that questioned the
validity of the HQE.
\begin{itemize}
\item In the mid-nineties there was the {\it missing charm puzzle} 
      (see e.g. \cite{Lenz:2000kv} for a brief review) - a disagreement between experiment and theory about the 
      average number of charm-quarks produced per $b$-decay. This issue has been resolved, by both improved 
      measurements and improved theory predictions \cite{Krinner:2013cja}.
\item For a long time the $\Lambda_b$ lifetime was measured to be considerably shorter than theoretically 
      expected, this issue has been resolved experimentally, mostly by the
      LHCb Collaboration (e.g. \cite{Aaij:2013oha,Aaij:2014owa,Aaij:2014zyy}) but also from the TeVatron 
      experiments \cite{Aaltonen:2014wfa}.
      The history of the {\it $\Lambda_b$-lifetime puzzle} and also attempts to obtain low theory values are
      discussed in detail in the
      review \cite{Lenz:2014jha}.
      The current status of lifetimes is depicted in Fig. \ref{lifetime}, taken from \cite{dordei}.
      \begin{figure}
      \includegraphics[width=\textwidth,angle=0]{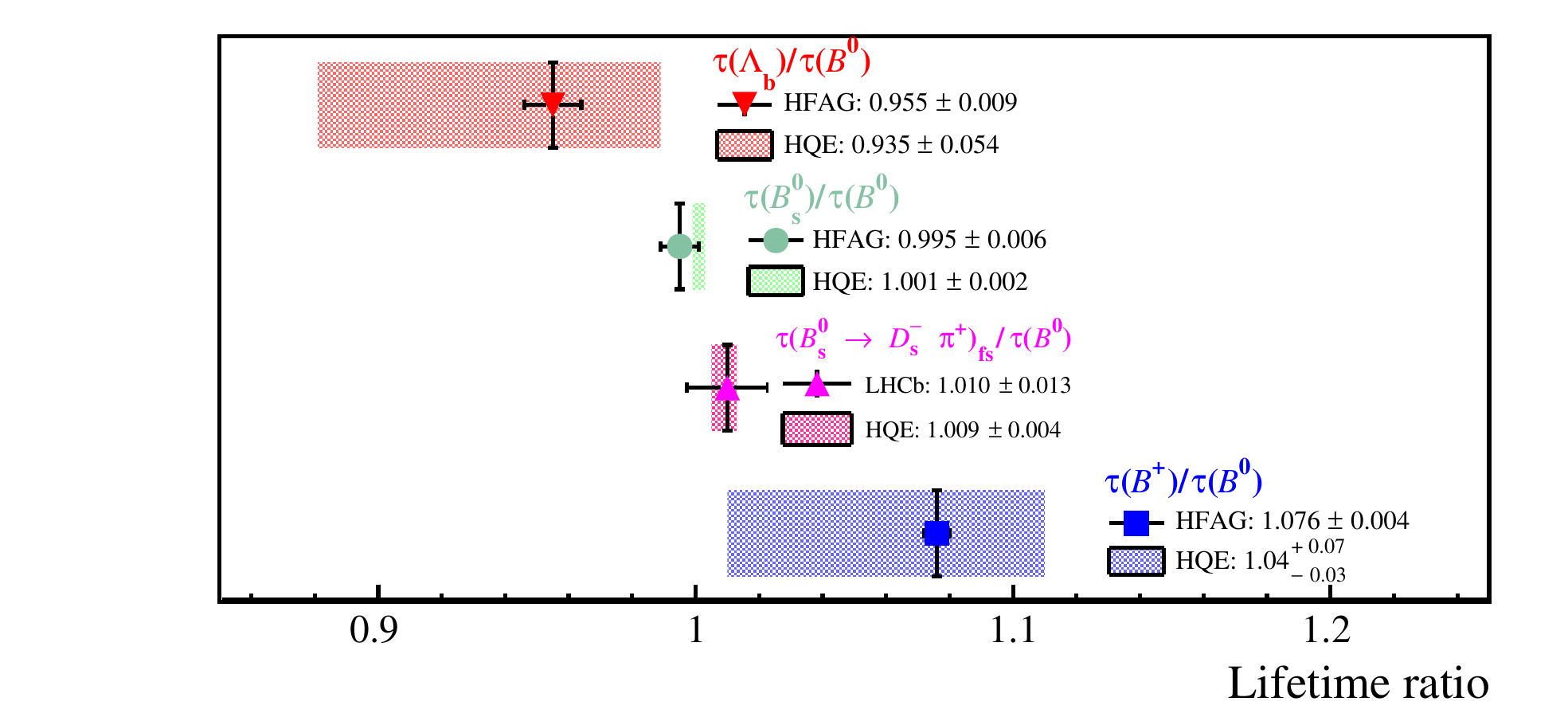}
      \caption{Comparison of HQE predictions for lifetime ratios of heavy hadrons with experimental values.
               The theory values are taken from \cite{Lenz:2014jha}, which is based on the calculations in
               \cite{Rosner:1996fy,Colangelo:1996ta,Baek:1998vk,DiPierro:1998ty,Cheng:1998ia,DiPierro:1999tb,Becirevic:2001fy,Beneke:2002rj,Franco:2002fc,Gabbiani:2004tp}. Experimental numbers are taken from HFAG \cite{HFAG}.
      The figure is taken from \cite{dordei}.}      
      \label{lifetime}
      \end{figure}
      One finds a nice agreement between experiment and theory and no lifetime puzzle exists anymore.
      The theoretical precision is, however, strongly limited by a lack of up-to-date values for the arising
      non-perturbative parameters. For the $\Lambda_b$-baryon the most recent lattice numbers stem from
      1999 \cite{DiPierro:1999tb} and for the $B$-mesons the most recent numbers are from 2001 
      \cite{Becirevic:2001fy}.
\item The applicability of the HQE was in particular questioned for $\Delta \Gamma_s$, see e.g.
      \cite{Lenz:2011zz}. In the last years this was also related to the unexpected measurement of a large value of the
      di-muon asymmetry by the D0 collaboration \cite{Abazov:2013uma,Abazov:2011yk,Abazov:2010hj,Abazov:2010hv}.
      The issue of $\Delta \Gamma_s$ was solved experimentally - mostly by the LHC experiments LHCb, ATLAS and 
      CMS and the current HFAG \cite{HFAG} average is in perfect agreement with the HQE prediction
      \cite{Lenz:2011ti} based on \cite{Lenz:2006hd,Beneke:2003az,Ciuchini:2003ww,Beneke:1998sy} - see \cite{Beneke:2000cu}
      for a very early prediction with NLO-QCD effects.
      \begin{eqnarray}
       \left( \frac{\Delta \Gamma_s}{\Delta M_s} \right)^{\rm Exp}
       / 
      \left( \frac{\Delta \Gamma_s}{\Delta M_s} \right)^{\rm SM}
      & = & 1.02 \pm 0.09 \pm 0.19 \; .
      \end{eqnarray}
      Again  an impressive confirmation of the HQE. The case of the di-muon asymmetry is still not settled yet.
      A new light was shed on it by the analysis of Borissov and Hoeneisen \cite{Borissov:2013wwa}, who
      found that the measured asymmetry does not only have contributions proportional to $a_{sl}^{d,s}$, but
      also some that originate from interference between mixing and decay and that might be approximated by 
      being proportional to $\Delta \Gamma_d$, see e.g. \cite{nierste} for a more detailed discussion.
\end{itemize}
All in all the HQE has been experimentally proven to be very successful and one could try to test its 
applicability also for charm-physics, see e.g. \cite{Lenz:2013aua,Bobrowski:2010xg} 
for some first investigations, or one can apply  the HQE now also to quantities that are sensitive 
to new physics, in particular to the semi-leptonic CP asymmetries. Their SM values are \cite{Lenz:2011ti}:
        \begin{eqnarray}
        a_{fs}^s =  \left(1.9  \pm 0.3 \right) \cdot 10^{-5} \; ,
        && 
        \phi_s   =  0.22^\circ \pm 0.06^\circ \; ,
        \\ 
        a_{fs}^d =  - \left(4.1 \pm 0.6 \right) \cdot 10^{-4} \; ,
        && 
        \phi_d  =  {-4.3^\circ} \pm 1.4^\circ \; .
        \end{eqnarray}
First measurements of these asymmetries \cite{Aaij:2013gta,Abazov:2012zz,Abazov:2012hha,Lees:2013sua}
are in agreement with the SM, but leave still some sizable space for NP effects.
\begin{eqnarray}
a_{sl}^{s \; \rm LHCb} =  -0.06 \pm 0.50 \pm 0.36 \% \; ,
&&
a_{sl}^{s \; \rm D0}   =  -1.12 \pm 0.74 \pm 0.17 \% \; ,
\\
a_{sl}^{d \; \rm D0}   =   \; \; \; 0.68 \pm 0.45 \pm 0.14  \% \; ,
&&
a_{sl}^{d \; \rm BaBar}=   \; \; \; 0.06 \pm 0.17 ^{+0.38}_{-0.32} \% \; .
\end{eqnarray}
At this workshop also some new preliminary numbers have been presented \cite{prel}
\begin{eqnarray}
a_{sl}^{d \; \rm LHCb} =  -0.02 \pm 0.19 \pm 0.30 \% \; ,
&&
a_{sl}^{d \; \rm BaBar}=  -0.39 \pm 0.35 \pm 0.19 \% \; .
\end{eqnarray}

\section{New physics effects in mixing}
\begin{figure}
      \includegraphics[width=0.8\textwidth,angle=0]{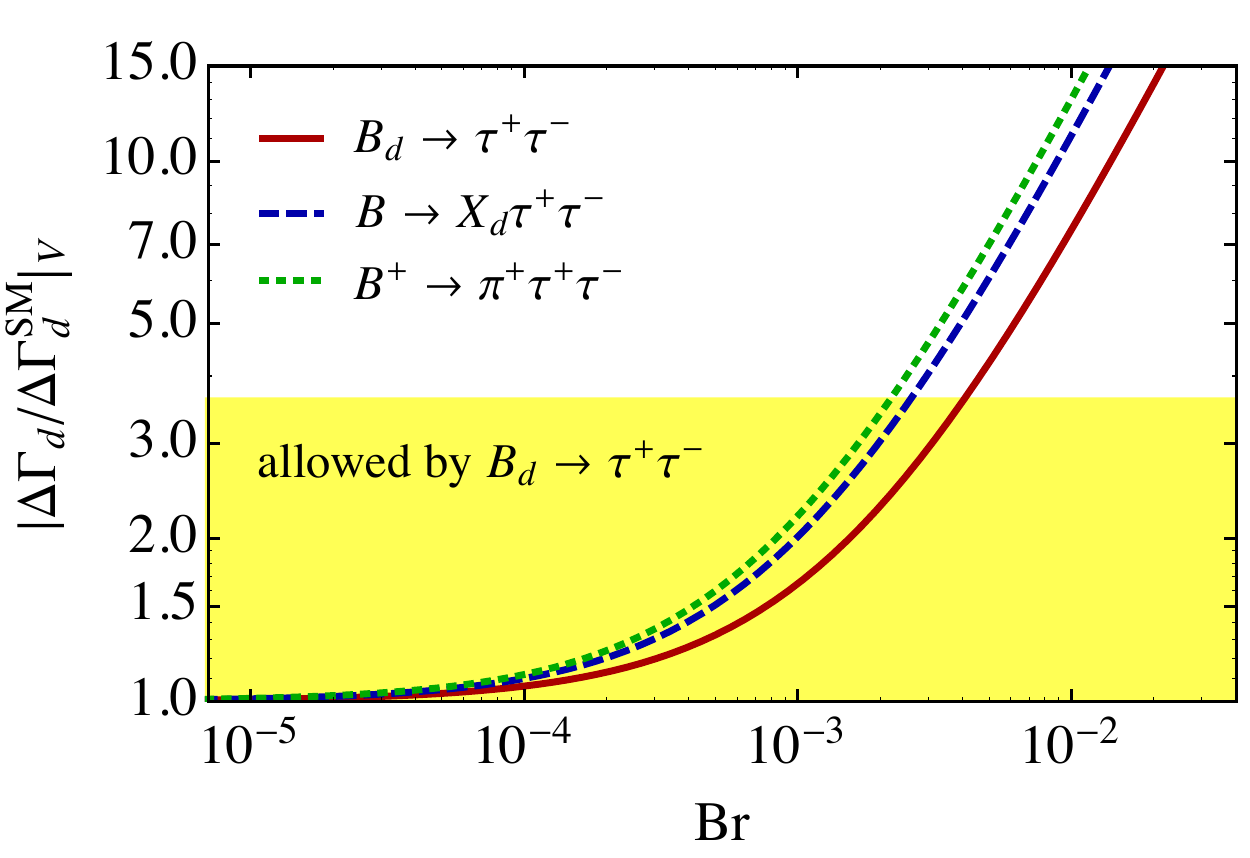}
      \caption{Required experimental precision in the decays $B_d \to \tau \tau$, $B \to X_d \tau \tau$
      and $B^+ \to \pi^+ \tau \tau$ in order to get a stronger bound on $\Delta \Gamma_d$ 
      than currently available (yellow region).}
      \label{tautau}
      \end{figure}

A reasonable start to search model-independently for new physics effects in $B$-mixing is the assumption that new physics 
only arises in $M_{12}^q$, i.e. $M_{12}^q = \Delta_q \cdot M_{12}^{q \; \rm SM}$ and  $\Gamma_{12}^q = \Gamma_{12}^{q \; \rm SM}$.
All new effects are encoded in the complex parameter $\Delta_q$. A corresponding strategy was suggested in \cite{Lenz:2006hd}
and worked out with real data in \cite{Lenz:2010gu,Lenz:2012az}. It turns out again  that everything is consistent 
with the SM and there are no huge NP effects, but there is still some space for sizable NP effects.
\\
This results also implies the necessity of a higher precision in our theory investigations and in 
particular it might be reasonable to take smaller NP effects in $\Delta \Gamma_q$ into account. 
In \cite{Bobeth:2011st} it was shown that for 
$\Delta \Gamma_s$ these effects can be at most of the order of $30 \%$, because else other experimental constraints 
will be violated. This is not the case for $\Delta \Gamma_d$, which has a very small SM value \cite{Lenz:2011ti}
and is only weakly constrained by measurements
       \begin{eqnarray}
        \left| \frac{\Delta \Gamma_d}{\Gamma_d}\right|^{\rm SM} = 
         \left(4.2 \pm 0.8\right) \cdot 10^{-3} \; ,
        &&
        \left| \frac{\Delta \Gamma_d}{\Gamma_d}\right|^{\rm HFAG} = 
         \left(1 \pm 10\right) \cdot 10^{-3} \; .
        \end{eqnarray}
In \cite{Bobeth:2014rda} three general scenarios were investigated in order to show that a enhancement of 
$\Delta \Gamma_d$ of several hundred per cent is currently not excluded. These were a violation of CKM unitarity,
new $bd \tau \tau$ operators and new physics effects on tree-level decays that act differently in the decays
$b \to c \bar{c} d$, $b \to c \bar{u} d$, $b \to u \bar{c} d$ and $b \to c \bar{c} d$ \footnote{Such non-universal, 
new tree-level effects can also affect the precision of the determination of the CKM angle $\gamma$.}.
Here first measurements in the $bd \tau \tau$ sector might yield some surprises, Fig.\ref{tautau}
shows the required experimental precision; stronger constraints on
the tree-level Wilson coefficients $C_1$ and $C_2$ would also be very helpful.

\section{Some very new physics effects}

Mixing of heavy mesons can also be used to test the fundamentals of quantum mechanics, see e.g.
\cite{Datta:1986ut,Bertlmann:1996at}. It was suggested to measure the ratio $R$ of like-sign dilepton
events and opposite-sign dilepton events and denote hypothetical deviations from the quantum mechanical coherence with the phenomenological parameter $\zeta$.
\begin{eqnarray}
R & = & \frac{N^{++} + N^{--}}{N^{+-} + N^{-+}} 
=
\frac12 \left( 
\left| \frac{p}{q} \right|^2
+
\left| \frac{q}{p} \right|^2
\right)
\frac{x^2+y^2       +{\zeta} \left[ y^2 \frac{1+x^2}{1-y^2} + x^2 \frac{1-y^2}{1+x^2} \right]}
     {2 + x^2 - y^2 +{\zeta} \left[ y^2 \frac{1+x^2}{1-y^2} - x^2 \frac{1-y^2}{1+x^2} \right]} \; .
\end{eqnarray}
Triggered by the 2013 paper of Alok and Banerjee \cite{Alok:2013sca}, which found 
extreme precise limits for decoherence effects, I redid the analysis with six talented 
undergraduate students \cite{Hodges:2014zaa} and we found a flaw in the arguments of 
\cite{Alok:2013sca}. Using the most recent values for  $x$ and $y$ from HFAG \cite{HFAG} and 
for $R$ from ARGUS \cite{Albrecht:1993gr} (1994) and CLEO \cite{Bartelt:1993cf} (1993)
we find that currently decoherence in $B$-mixing is only very loosely bounded 
\begin{equation}
\zeta = -0.26^{+0.30}_{-0.28} \; .
\end{equation}
Here future measurements would be very helpful to gain additional insights. 
To demonstrate the required experimental precision in $R$, we show how the error
in $R$ affects the uncertainty in $\zeta$.
\begin{equation}
\begin{array}{|c||c|c|c|}
\hline
\delta R & \pm 10 \%  & \pm 5 \%  & \pm 2 \% 
\\
\hline
\delta \zeta & ^{+45.2\%}_{-43.8 \%} & 
               ^{+22.8 \%}_{-22.4 \%} & ^{+10.0\%}_{-9.98 \%} 
\\
\hline
\end{array}
\; .
\end{equation}

\section{Conclusion}
The HQE has been successfully tested by many recent experiments, further more precise
tests of the HQE demand non-perturbative input, mostly matrix elements of dimension six
operators. Applying the HQE predictions to NP sensitive quantities one finds that everything
is consistent with the SM, but there is still some space left for new effects. Promising observables in that
respect are more precise values of $a_{sl}^{d,s}$ and $\Delta \Gamma_d$, 
first measurements of $bd \tau \tau$ and  $bs \tau \tau$-transitions as well as further constraints
on the tree-level Wilson coefficients $C_{1,2}$. Finally we suggest also a new measurement of the 
ratio $R$ of like-sign dilepton
events and opposite-sign dilepton events.

\Acknowledgements
I would like to thank the organisers for creating such a pleasant scientific event.


\begin{thebibliography}{99}

\bibitem{Anikeev:2001rk}
  K.~Anikeev, D.~Atwood, F.~Azfar, S.~Bailey, C.~W.~Bauer, W.~Bell, G.~Bodwin and E.~Braaten {\it et al.},
  hep-ph/0201071.


\bibitem{Bediaga:2012py}
  R.~Aaij {\it et al.}  [LHCb Collaboration],
  Eur.\ Phys.\ J.\ C {\bf 73} (2013) 2373
  [arXiv:1208.3355 [hep-ex]].

\bibitem{Lenz:2014nka}
  A.~J.~Lenz,
  J.\ Phys.\ G {\bf 41} (2014) 103001
  [arXiv:1404.6197 [hep-ph]].

\bibitem{Lenz:2012mb}
  A.~Lenz,
  arXiv:1205.1444 [hep-ph].


\bibitem{kubo}
 Y.~Kaburaki, K.~Konya, J.~Kubo and A.~Lenz,
  Phys.\ Rev.\ D {\bf 84} (2011) 016007
  [arXiv:1012.2435 [hep-ph]];
 J.~Kubo and A.~Lenz,
  Phys.\ Rev.\ D {\bf 82} (2010) 075001
  [arXiv:1007.0680 [hep-ph]];
 K.~Kawashima, J.~Kubo and A.~Lenz,
  Phys.\ Lett.\ B {\bf 681} (2009) 60
  [arXiv:0907.2302 [hep-ph]];
N.~Kifune, J.~Kubo and A.~Lenz,
  Phys.\ Rev.\ D {\bf 77} (2008) 076010
  [arXiv:0712.0503 [hep-ph]].

\bibitem{Inami:1980fz}
  T.~Inami and C.~S.~Lim,
  Prog.\ Theor.\ Phys.\  {\bf 65} (1981) 297
   [Erratum-ibid.\  {\bf 65} (1981) 1772].

\bibitem{Buras:1990fn}
  A.~J.~Buras, M.~Jamin and P.~H.~Weisz,
  Nucl.\ Phys.\ B {\bf 347} (1990) 491.

\bibitem{Aoki:2013ldr}
  S.~Aoki, Y.~Aoki, C.~Bernard, T.~Blum, G.~Colangelo, M.~Della Morte, S.~Dürr and A.~X.~El Khadra {\it et al.},
  arXiv:1310.8555 [hep-lat].
  \\
  and online update at http://itpwiki.unibe.ch/flag


\bibitem{HFAG}
Y.~Amhis {\it et al.}  [Heavy Flavor Averaging Group Collaboration],
  arXiv:1207.1158 [hep-ex].
  \\
   online update at http://www.slac.stanford.edu/xorg/hfag

\bibitem{Bazavov:2011aa}
  A.~Bazavov {\it et al.}  [Fermilab Lattice and MILC Collaborations],
  Phys.\ Rev.\ D {\bf 85} (2012) 114506
  [arXiv:1112.3051 [hep-lat]].


\bibitem{Dowdall:2013tga}
  R.~J.~Dowdall {\it et al.}  [HPQCD Collaboration],
  Phys.\ Rev.\ Lett.\  {\bf 110} (2013) 22,  222003
  [arXiv:1302.2644 [hep-lat]].




\bibitem{Lenz:2014jha}
  A.~Lenz,
  arXiv:1405.3601 [hep-ph].

\bibitem{Lenz:2000kv}
  A.~Lenz,
  hep-ph/0011258.

\bibitem{Krinner:2013cja}
  F.~Krinner, A.~Lenz and T.~Rauh,
  Nucl.\ Phys.\ B {\bf 876} (2013) 31
  [arXiv:1305.5390 [hep-ph]].


\bibitem{Aaij:2013oha}
  R.~Aaij {\it et al.}  [LHCb Collaboration],
  Phys.\ Rev.\ Lett.\  {\bf 111} (2013) 102003
  [arXiv:1307.2476 [hep-ex]].

\bibitem{Aaij:2014owa}
  R.~Aaij {\it et al.}  [LHCb Collaboration],
  JHEP {\bf 1404} (2014) 114
  [arXiv:1402.2554 [hep-ex], arXiv:1402.2554].


\bibitem{Aaij:2014zyy}
  R.~Aaij {\it et al.}  [LHCb Collaboration],
  Phys.\ Lett.\ B {\bf 734} (2014) 122
  [arXiv:1402.6242 [hep-ex]].


\bibitem{Aaltonen:2014wfa}
  T.~A.~Aaltonen {\it et al.}  [CDF Collaboration],
  Phys.\ Rev.\ D {\bf 89} (2014) 072014
  [arXiv:1403.8126 [hep-ex]].



\bibitem{dordei}
Courtesy of Francesca Dordei \\  talk at CKM 2014, Vienna , https://cds.cern.ch/record/1756840.




\bibitem{Rosner:1996fy}
  J.~L.~Rosner,
  Phys.\ Lett.\ B {\bf 379} (1996) 267
  [hep-ph/9602265].

\bibitem{Colangelo:1996ta}
  P.~Colangelo and F.~De Fazio,
  Phys.\ Lett.\ B {\bf 387} (1996) 371
  [hep-ph/9604425].

\bibitem{Baek:1998vk}
  M.~S.~Baek, J.~Lee, C.~Liu and H.~S.~Song,
  Phys.\ Rev.\ D {\bf 57} (1998) 4091
  [hep-ph/9709386].

\bibitem{DiPierro:1998ty}
  M.~Di Pierro {\it et al.}  [UKQCD Collaboration],
  Nucl.\ Phys.\ B {\bf 534} (1998) 373
  [hep-lat/9805028].

\bibitem{Cheng:1998ia}
  H.~Y.~Cheng and K.~C.~Yang,
  Phys.\ Rev.\ D {\bf 59} (1999) 014011
  [hep-ph/9805222].

\bibitem{DiPierro:1999tb}
  M.~Di Pierro {\it et al.}  [UKQCD Collaboration],
  Phys.\ Lett.\ B {\bf 468} (1999) 143
  [hep-lat/9906031].

\bibitem{Becirevic:2001fy}
  D.~Becirevic,
  PoS HEP {\bf 2001} (2001) 098
  [hep-ph/0110124].

\bibitem{Beneke:2002rj}
  M.~Beneke, G.~Buchalla, C.~Greub, A.~Lenz and U.~Nierste,
  Nucl.\ Phys.\ B {\bf 639} (2002) 389
  [hep-ph/0202106].

\bibitem{Franco:2002fc}
  E.~Franco, V.~Lubicz, F.~Mescia and C.~Tarantino,
  Nucl.\ Phys.\ B {\bf 633} (2002) 212
  [hep-ph/0203089].


\bibitem{Gabbiani:2004tp}
  F.~Gabbiani, A.~I.~Onishchenko and A.~A.~Petrov,
  Phys.\ Rev.\ D {\bf 70} (2004) 094031
  [hep-ph/0407004].


\bibitem{Lenz:2011zz}
  A.~J.~Lenz,
  Phys.\ Rev.\ D {\bf 84} (2011) 031501
  [arXiv:1106.3200 [hep-ph]].


\bibitem{Abazov:2013uma}
  V.~M.~Abazov {\it et al.}  [D0 Collaboration],
  Phys.\ Rev.\ D {\bf 89} (2014) 1,  012002
  [arXiv:1310.0447 [hep-ex]].

\bibitem{Abazov:2011yk}
  V.~M.~Abazov {\it et al.}  [D0 Collaboration],
  Phys.\ Rev.\ D {\bf 84} (2011) 052007
  [arXiv:1106.6308 [hep-ex]].

\bibitem{Abazov:2010hj}
  V.~M.~Abazov {\it et al.}  [D0 Collaboration],
  Phys.\ Rev.\ Lett.\  {\bf 105} (2010) 081801
  [arXiv:1007.0395 [hep-ex]].

\bibitem{Abazov:2010hv}
  V.~M.~Abazov {\it et al.}  [D0 Collaboration],
  Phys.\ Rev.\ D {\bf 82} (2010) 032001
  [arXiv:1005.2757 [hep-ex]].


\bibitem{Lenz:2011ti}
  A.~Lenz and U.~Nierste,
  arXiv:1102.4274 [hep-ph].

\bibitem{Lenz:2006hd}
  A.~Lenz and U.~Nierste,
  JHEP {\bf 0706} (2007) 072
  [hep-ph/0612167].

\bibitem{Beneke:2003az}
  M.~Beneke, G.~Buchalla, A.~Lenz and U.~Nierste,
  Phys.\ Lett.\ B {\bf 576} (2003) 173
  [hep-ph/0307344].


\bibitem{Ciuchini:2003ww}
  M.~Ciuchini, E.~Franco, V.~Lubicz, F.~Mescia and C.~Tarantino,
  JHEP {\bf 0308} (2003) 031
  [hep-ph/0308029].

\bibitem{Beneke:1998sy}
  M.~Beneke, G.~Buchalla, C.~Greub, A.~Lenz and U.~Nierste,
  Phys.\ Lett.\ B {\bf 459} (1999) 631
  [hep-ph/9808385].

\bibitem{Beneke:2000cu}
  M.~Beneke and A.~Lenz,
  J.\ Phys.\ G {\bf 27} (2001) 1219
  [hep-ph/0012222].


\bibitem{Borissov:2013wwa}
  G.~Borissov and B.~Hoeneisen,
  Phys.\ Rev.\ D {\bf 87} (2013) 7,  074020
  [arXiv:1303.0175 [hep-ex]].

\bibitem{nierste}
U. Nierste, these proceedings.

\bibitem{Lenz:2013aua}
  A.~Lenz and T.~Rauh,
  Phys.\ Rev.\ D {\bf 88} (2013) 034004
  [arXiv:1305.3588 [hep-ph]].

\bibitem{Bobrowski:2010xg}
  M.~Bobrowski, A.~Lenz, J.~Riedl and J.~Rohrwild,
  JHEP {\bf 1003} (2010) 009
  [arXiv:1002.4794 [hep-ph]].

\bibitem{Aaij:2013gta}
  R.~Aaij {\it et al.}  [LHCb Collaboration],
  Phys.\ Lett.\ B {\bf 728} (2014) 607
  [arXiv:1308.1048 [hep-ex]].

\bibitem{Abazov:2012zz} 
  V.~M.~Abazov {\it et al.}  [D0 Collaboration],
  Phys.\ Rev.\ Lett.\  {\bf 110}, 011801 (2013)
  [arXiv:1207.1769 [hep-ex]].


\bibitem{Abazov:2012hha}
  V.~M.~Abazov {\it et al.}  [D0 Collaboration],
  Phys.\ Rev.\ D {\bf 86} (2012) 072009
  [arXiv:1208.5813 [hep-ex]].

\bibitem{Lees:2013sua}
  J.~P.~Lees {\it et al.}  [BaBar Collaboration],
  Phys.\ Rev.\ Lett.\  {\bf 111} (2013) 10,  101802
   [Addendum-ibid.\  {\bf 111} (2013) 15,  159901]
  [arXiv:1305.1575 [hep-ex]].

\bibitem{prel}
L. Grillo, these proceedings,\\
C. Cheng, these proceedings.

\bibitem{Lenz:2010gu}
  A.~Lenz, U.~Nierste, J.~Charles, S.~Descotes-Genon, A.~Jantsch, C.~Kaufhold, H.~Lacker and S.~Monteil {\it et al.},
  Phys.\ Rev.\ D {\bf 83} (2011) 036004
  [arXiv:1008.1593 [hep-ph]].

\bibitem{Lenz:2012az}
  A.~Lenz, U.~Nierste, J.~Charles, S.~Descotes-Genon, H.~Lacker, S.~Monteil, V.~Niess and S.~T'Jampens,
  Phys.\ Rev.\ D {\bf 86} (2012) 033008
  [arXiv:1203.0238 [hep-ph]].


\bibitem{Bobeth:2011st}
  C.~Bobeth and U.~Haisch,
  Acta Phys.\ Polon.\ B {\bf 44} (2013) 127
  [arXiv:1109.1826 [hep-ph]].


\bibitem{Bobeth:2014rda}
  C.~Bobeth, U.~Haisch, A.~Lenz, B.~Pecjak and G.~Tetlalmatzi-Xolocotzi,
  arXiv:1404.2531 [hep-ph].




\bibitem{Datta:1986ut}
  A.~Datta and D.~Home,
  Phys.\ Lett.\ A {\bf 119} (1986) 3.

\bibitem{Bertlmann:1996at}
  R.~A.~Bertlmann and W.~Grimus,
  Phys.\ Lett.\ B {\bf 392} (1997) 426
  [hep-ph/9610301].

\bibitem{Alok:2013sca}
  A.~K.~Alok and S.~Banerjee,
  Phys.\ Rev.\ D {\bf 88} (2013) 9,  094013
  [arXiv:1304.4063 [hep-ph]].

\bibitem{Hodges:2014zaa}
  D.~Hodges, D.~Hulme, S.~Kvedaraite, A.~Lenz, J.~Richings, J.~S.~Woo and P.~Waite,
  arXiv:1408.0222 [hep-ph].

\bibitem{Albrecht:1993gr}
  H.~Albrecht {\it et al.}  [ARGUS Collaboration],
  Phys.\ Lett.\ B {\bf 324} (1994) 249.


\bibitem{Bartelt:1993cf}
  J.~E.~Bartelt {\it et al.}  [CLEO Collaboration],
  Phys.\ Rev.\ Lett.\  {\bf 71} (1993) 1680.


\end{thebibliography}
\end{document}